\documentclass[11pt]{article}
\begin{document}
\title{{\bf{\Large Fermion Tunneling Beyond Semiclassical Approximation}}}
\author{
 {\bf {\normalsize Bibhas Ranjan Majhi}$
$\thanks{E-mail: bibhas@bose.res.in}}\\
 {\normalsize S.~N.~Bose National Centre for Basic Sciences,}
\\{\normalsize JD Block, Sector III, Salt Lake, Kolkata-700098, India}
\\[0.3cm]
}

\maketitle

{\bf Abstract:}\\
        Applying the Hamilton-Jacobi method beyond the semiclassical approximation prescribed in \cite{Majhi3} for the scalar particle, Hawking radiation as tunneling of Dirac particle through an event horizon is analysed. We show that, as before, all quantum corrections in the single particle action are proportional to the usual semiclassical contribution. We also compute the modifications to the Hawking temperature and Bekenstein-Hawking entropy for the Schwarzschild black hole. Finally, the coefficient of the logarithmic correction to entropy is shown to be related with the trace anomaly.\\\\

        Semiclassical methods of modeling Hawking radiation as a tunneling effect were developed over the past decade and have generated a lot of interest \cite{Wilczek,Paddy,Berezin,Medved1,Jiang,Singleton,Vanzo,Majhi1,Majhi2,Majhi5}. From this approach an alternative (intuitive) way of understanding black hole radiation emerged. However, most of the calculations in the literature \cite{Wilczek,Paddy,Medved1,Jiang,Singleton,Vanzo,Majhi1,Majhi2,Majhi5} have been performed just for scalar particles. Since a black hole can radiate all types of particles like a black body, the emission spectrum should contain particles of all spins. Therefore a detailed study of spin one-half particle emission is necessary. Although there exist some computations \cite{Chen} in this context, these are confined to the semiclassical approximation and do not consider quantum corrections.

       In our previous work \cite{Majhi3}, we formulated the Hamilton-Jacobi method of tunneling beyond semiclassical approximation by considering all the terms in the expansion of the one particle action for a scalar particle. We showed that the higher order terms are proportional to the semiclassical contribution. This result, together with properties of conformal transformations, eventually led to corrected expressions for thermodynamic variables of a black hole. It is not obvious whether a similar analysis is valid for the case of spin-half fermion tunneling. This issue is addressed here.

       In this paper we will discuss the Dirac particle tunneling beyond semiclassical approximation employing the Hamilton-Jacobi method suggested in \cite{Majhi3}. We will explicitly show that the higher order terms in the single particle action are again proportional to the semiclassical contribution. By dimensional argument the form of these proportionality constants, upto some dimensionless parameters, are determined. In particular for Scwarzschild spacetime, these are given by the inverse powers of the square of the mass of the black hole, because in this case, the only macroscopic parameter is mass. Using the principle of ``detailed balance'' \cite{Paddy,Majhi2} the modified Hawking temperature is identified. Then the corrections to the Bekenstein-Hawking area law are derived by using the Gibbs form of first law of thermodynamics. Interestingly, the leading order correction to the entropy is the logarithmic of the semiclassical entropy which was found earlier in \cite{Fursaev,Mann,Majumdar,Suneeta,Das,More,Sudipta,Mitra,Page,Modak}. Finally, using a constant scale transformation to the metric, we show that the coefficient of the logarithmic correction is related to trace anomaly.\\

     Our method involves calculating the imaginary part of the action for the (classically forbidden) process of s-wave emission across the horizon which in turn is related to the Boltzmann factor for emission at the Hawking temperature. We consider a massless Dirac particle in a general class of static, spherically symmetric spacetime of the form
\begin{eqnarray}
ds^2=-f(r)dt^2+\frac{dr^2}{g(r)}+r^2d\Omega^2
\label{1.01}
\end{eqnarray}
where the horizon $r=r_H$ is given by $f(r_H)=g(r_H)=0$. The massless Dirac equation is given by
\begin{eqnarray}
i\gamma^\mu\nabla_\mu\psi=0
\label{1.02}
\end{eqnarray}
where for this case the $\gamma$ matrices are defined as,
\begin{eqnarray}
\gamma^t&=&\frac{1}{\sqrt{f(r)}}\left(\begin{array}{c c}
i & 0 \\
0 & -i
\end{array}\right);\,\,\
\gamma^r=\sqrt{g(r)}\left(\begin{array}{c c}
0 & \sigma^3 \\
\sigma^3 & 0
\end{array}\right)
\nonumber
\\
\gamma^\theta&=&\frac{1}{r}\left(\begin{array}{c c}
0 & \sigma^1 \\
\sigma^1 & 0
\end{array}\right);\,\,\,\
\gamma^\phi=\frac{1}{r\textrm{sin}\theta}\left(\begin{array}{c c}
0 & \sigma^2 \\
\sigma^2 & 0
\end{array}\right).
\label{1.03}
\end{eqnarray}
The covariant derivative is given by,
\begin{eqnarray}
&&\nabla_\mu=\partial_\mu+\frac{i}{2}\Gamma{^\alpha}{_\mu}{^\beta}\Sigma_{\alpha\beta};\,\,\ \Gamma{^\alpha}{_\mu}{^\beta}=g^{\beta\nu}\Gamma^\alpha_{\mu\nu};\,\,\ \Sigma_{\alpha\beta}=\frac{i}{4}\Big[\gamma_\alpha,\gamma_\beta\Big]
\nonumber
\\
&&\{\gamma^\mu,\gamma^\nu\}=2g^{\mu\nu}
\label{1.04}
\end{eqnarray}
Since for radial trajectories only the $(r-t)$ sector of the metric (\ref{1.01}) is important, (\ref{1.02}) can be expressed as
\begin{eqnarray}
i\gamma^\mu\partial_\mu\psi-\frac{1}{2}\Big(g^{tt}\gamma^\mu\Gamma^r_{\mu t}-g^{rr}\gamma^\mu\Gamma^t_{\mu r}\Big)\Sigma_{rt}\psi=0.
\label{1.05}
\end{eqnarray}
Here the required nonvanishing connections are
\begin{eqnarray}
\Gamma^r_{tt}=\frac{f'g}{2};\,\,\ \Gamma^t_{tr}=\frac{f'}{2f}.
\label{1.06}
\end{eqnarray}
Therefore, under the metric (\ref{1.01}), the Dirac equation (\ref{1.02}) reduces to
\begin{eqnarray}
i\gamma^t\partial_t\psi+i\gamma^r\partial_r\psi+\frac{f'g}{2f}\gamma^t\Sigma_{rt}\psi=0
\label{1.07}
\end{eqnarray}
and the matrix form of $\Sigma_{rt}$ from (\ref{1.03}) and (\ref{1.04})  is given by
\begin{eqnarray}
\Sigma_{rt}=\frac{i}{2}\left(\begin{array}{c c c c}
0 & 0 & i\sqrt{\frac{f}{g}} & 0\\ 
0 & 0 & 0 & -i\sqrt{\frac{f}{g}}\\
-i\sqrt{\frac{f}{g}} & 0 & 0 & 0\\
0 & i\sqrt{\frac{f}{g}} & 0 & 0
\end{array}\right).
\label{1.08}
\end{eqnarray}
To solve (\ref{1.07}) we employ the following ansatz for the spin up (i.e. +ve $r$-direction) and spin down (i.e. -ve $r$-direction) $\psi$ as
\begin{eqnarray}
\psi_\uparrow(t,r)= \left(\begin{array}{c}
A(t,r) \\
0 \\
B(t,r) \\
0
\end{array}\right){\textrm{exp}}\Big[\frac{i}{\hbar}I_\uparrow (t,r)\Big]
\label{1.09}
\end{eqnarray}  
\begin{eqnarray}
\psi_\downarrow(t,r)= \left(\begin{array}{c}
0 \\
C(t,r) \\
0 \\
D(t,r) \\
\end{array}\right){\textrm{exp}}\Big[\frac{i}{\hbar}I_\downarrow (t,r)\Big]
\label{1.10}
\end{eqnarray}
where $I(r,t)$ is the one particle action which will be expanded in powers of $\hbar$. Here we will only solve the spin up case explicitly since the spin down case is fully analogous.  Substituting the ansatz (\ref{1.09}) in (\ref{1.07}), we obtain the following two equations:
\begin{eqnarray}
\Big(\frac{iA}{\sqrt{f}}\partial_t I_\uparrow + B\sqrt{g}\partial_r I_\uparrow\Big)+\hbar\Big(\frac{1}{\sqrt{f}}\partial_t A-i\sqrt{g}\partial_r B+i\frac{f'\sqrt{g}}{4f}B\Big)=0
\label{1.11}
\end{eqnarray}
\begin{eqnarray}
\Big(-\frac{iB}{\sqrt{f}}\partial_t I_\uparrow + A\sqrt{g}\partial_r I_\uparrow\Big)+\hbar\Big(-\frac{1}{\sqrt{f}}\partial_t B-i\sqrt{g}\partial_r A+i\frac{f'\sqrt{g}}{4f}A\Big)=0.
\label{1.12}
\end{eqnarray}
Since the last terms within the first bracket of the above equations do not involve the single particle action, they will not contribute to the thermodynamic entities of the black hole. Therefore we will drop these two terms. Now taking $I_\uparrow=I$ and expanding $I, A$ and $B$ in powers of $\hbar$, we find,
\begin{eqnarray}
&&I(r,t)=I_0(r,t)+\displaystyle\sum_i \hbar^i I_i(r,t)
\nonumber
\\
&&A=A_0+\displaystyle\sum_i \hbar^i A_i; \,\,\, B=B_0+\displaystyle\sum_i \hbar^i B_i.
\label{1.13}
\end{eqnarray}
where $i=1,2,3,......$. In these expansions the terms from ${\cal{O}}(\hbar)$ onwards are treated as quantum corrections over the semiclassical value $I_0$, $A_0$ and $B_0$ respectively. Substituting (\ref{1.13}) in (\ref{1.11}) and (\ref{1.12}) and then equating the different powers of $\hbar$ on both sides, we obtain the following two sets of equations:
\begin{eqnarray}
{\textrm{Set I}}:~
\hbar^0:&&\frac{i}{\sqrt{f}}A_0\partial_t I_0+\sqrt{g}B_0\partial_r I_0=0
\label{ref1}
\\
\hbar^1:&&\frac{i}{\sqrt{f}}A_0\partial_t I_1+\frac{i}{\sqrt{f}}A_1\partial_t I_0+\sqrt{g}B_0\partial_r I_1+\sqrt{g}B_1\partial_r I_0=0
\label{ref2}
\\
\hbar^2:&&\frac{i}{\sqrt{f}}A_0\partial_t I_2+\frac{i}{\sqrt{f}}A_1\partial_t I_1+\frac{i}{\sqrt{f}}A_2\partial_t I_0
\nonumber
\\
&+&\sqrt{g}B_0\partial_r I_2+\sqrt{g}B_1\partial_r I_1+\sqrt{g}B_2\partial_r I_0=0
\label{ref3}
\\
&&.
\nonumber
\\
&&.
\nonumber
\\
&&.
\nonumber
\\
{\textrm{and so on.}}
\nonumber
\end{eqnarray}
\begin{eqnarray}
{\textrm{Set II}}:~
\hbar^0:&&-\frac{i}{\sqrt{f}}B_0\partial_t I_0+\sqrt{g}A_0\partial_r I_0=0
\label{ref4}
\\
\hbar^1:&&-\frac{i}{\sqrt{f}}B_0\partial_t I_1-\frac{i}{\sqrt{f}}B_1\partial_t I_0+\sqrt{g}A_0\partial_r I_1+\sqrt{g}A_1\partial_r I_0=0
\label{ref5}
\\
\hbar^2:&&-\frac{i}{\sqrt{f}}B_0\partial_t I_2-\frac{i}{\sqrt{f}}B_1\partial_t I_1-\frac{i}{\sqrt{f}}B_2\partial_t I_0
\nonumber
\\
&+&\sqrt{g}A_0\partial_r I_2+\sqrt{g}A_1\partial_r I_1+\sqrt{g}A_2\partial_r I_0=0
\label{ref6}
\\
&&.
\nonumber
\\
&&.
\nonumber
\\
&&.
\nonumber
\\
{\textrm{and so on.}}
\nonumber
\end{eqnarray}
Equations (\ref{ref1}) and (\ref{ref4}) are collectively known as the semiclassical Hamilton-Jacobi equations for a Dirac particle. Since the metric (\ref{1.01}) is stationary it has timelike Killing vectors. Thus we will look for solutions of (\ref{ref1}) and (\ref{ref4}) which behave as 
\begin{eqnarray}
I_0=\omega t + W(r),
\label{1.19}
\end{eqnarray} 
where $\omega$ is the energy of the particle. Substituting this in (\ref{ref1}) and (\ref{ref4}) we obtain,
\begin{eqnarray}
\frac{iA_0}{\sqrt{f}}\omega + B_0\sqrt{g}W'(r)=0
\nonumber
\\
-\frac{iB_0}{\sqrt{f}}\omega + A_0\sqrt{g}W'(r)=0.
\label{1.20}
\end{eqnarray}
These two equations have two possible solutions: 
\begin{eqnarray}
&&A_0=iB_0;\,\,\ W_+(r) = \omega\int_0^r \frac{dr}{\sqrt{f(r)g(r)}}
\nonumber
\\
&&A_0=-iB_0;\,\,\, W_-(r) =  -\omega\int_0^r\frac{dr}{\sqrt{f(r)g(r)}} 
\label{1.21}
\end{eqnarray}
where $W_+$($W_-$) corresponds to ingoing (outgoing) solutions. The limits of the integration are chosen such that the particle goes through the horizon $r=r_H$. Therefore the solution for $I_0(r,t)$ is 
\begin{eqnarray}
I_0(r,t)=\omega t\pm\omega\int_0^r\frac{dr}{\sqrt{f(r)g(r)}}.
\label{ref7}
\end{eqnarray}

   Now, it is interesting to note that using (\ref{1.21}) and (\ref{ref7}) in the equations of Set I and Set II simultaneously and then solving we get relations connecting different orders in the expansion of $A$ with those of $B$:
\begin{eqnarray}
A_a=\pm iB_a
\label{ref8}
\end{eqnarray}
where $a=0,1,2,3,....$. These lead to a simplified form of all the equations in Set I and Set II as,
\begin{eqnarray}
\partial_tI_a=\pm\sqrt{fg}\partial_rI_a
\label{1.14}
\end{eqnarray}
i.e. the functional form of the above individual linear differential equations is same and is identical to the usual semiclassical Hamilton-Jacobi equations (\ref{ref1}) and (\ref{ref4}). Therefore the solutions of these equations are not independent and $I_i$'s are proportional to $I_0$. A similar situation happened for scalar particle tunneling \cite{Majhi3}. Since $I_0$ has the dimension of $\hbar$ the proportionality constants should have the dimension of inverse of $\hbar^i$. Again in the units $G=c=k_B=1$ the Planck constant $\hbar$ is of the order of square of the Planck Mass $M_P$ and so from dimensional analysis the proportionality constants have the dimension of $M^{-2i}$ where $M$ is the mass of black hole. Specifically, for Schwarzschild type black holes having mass as the only macroscopic parameter, these considerations show that the most general expression for $I$, following from (\ref{1.13}), valid for (\ref{1.14}), is given by,
\begin{eqnarray}
I(r,t)=\Big(1+\sum_i\beta_i\frac{\hbar^i}{M^{2i}}\Big)I_0(r,t).
\label{1.16}
\end{eqnarray}
where $\beta_i$'s are dimensionless constant parameters.

   The above analysis shows that to obtain a solution for $I(r,t)$ it is therefore enough to solve for $I_0(r,t)$ which has the solution of the form (\ref{ref7}). 
In fact the standard Hamilton-Jacobi solution determined by this $I_0(r,t)$ is just modified by a prefactor to yield the complete solution for $I(r,t)$. Substituting (\ref{ref7}) in (\ref{1.16}) we obtain
\begin{eqnarray}
I(r,t)= \Big(1+\sum_i\beta_i\frac{\hbar^i}{M^{2i}}\Big)\Big[\omega t  \pm\omega\int_0^r\frac{dr}{\sqrt{f(r)g(r)}}\Big].
\label{1.22}
\end{eqnarray}  
Therefore the ingoing and outgoing solutions of the Dirac equation (\ref{1.02}) under the background metric (\ref{1.01}) is given by exploiting (\ref{1.09}) and (\ref{1.22}),
\begin{eqnarray}
\psi_{{\textrm {in}}}\sim {\textrm{exp}}\Big[-\frac{i}{\hbar}(1+\sum_i\beta_i\frac{\hbar^i}{M^{2i}})\Big(\omega t  +\omega\int_0^r\frac{dr}{\sqrt{f(r)g(r)}}\Big)\Big]
\label{1.23}
\end{eqnarray} 
and
\begin{eqnarray}
\psi_{{\textrm {out}}}\sim {\textrm{exp}}\Big[-\frac{i}{\hbar}(1+\sum_i\beta_i\frac{\hbar^i}{M^{2i}})\Big(\omega t  -\omega\int_0^r\frac{dr}{\sqrt{f(r)g(r)}}\Big)\Big].
\label{1.24}
\end{eqnarray} 
Now for the tunneling of a particle across the horizon the nature of the coordinates change. The sign of the metric coefficients in the $(r-t)$ sector is altered. This indicates that `$t$' coordinate has an imaginary part for the crossing of the horizon of the black hole and correspondingly there will be a temporal contribution to the probabilities for the ingoing and outgoing particles. This has similarity with\cite{Akhmedov} where they show for the Schwarzschild metric that two patches across the horizon are connected by a discrete imaginary amount of time.

     The ingoing and outgoing probabilities of the particle are, therefore, given by,
\begin{eqnarray}
P_{{\textrm{in}}}=|\psi_{{\textrm {in}}}|^2\sim {\textrm{exp}}\Big[\frac{2}{\hbar}(1+\sum_i\beta_i\frac{\hbar^i}{M^{2i}})\Big(\omega{\textrm{Im}}~t +\omega{\textrm{Im}}\int_0^r\frac{dr}{\sqrt{f(r)g(r)}}\Big)\Big]
\label{1.25}
\end{eqnarray}
and
\begin{eqnarray}
P_{{\textrm{out}}}=|\psi_{{\textrm {out}}}|^2\sim {\textrm{exp}}\Big[\frac{2}{\hbar}(1+\sum_i\beta_i\frac{\hbar^i}{M^{2i}})\Big(\omega{\textrm{Im}}~t -\omega{\textrm{Im}}\int_0^r\frac{dr}{\sqrt{f(r)g(r)}}\Big)\Big]
\label{1.26}
\end{eqnarray}
Now the ingoing probability $P_{\textrm{in}}$ has to be unity in the classical limit (i.e. $\hbar\rightarrow 0$) - when there is no reflection and everything is absorbed - instead of zero or infinity \cite{Majhi3}.Thus, in the classical limit, (\ref{1.25}) leads to,
\begin{eqnarray}
{\textrm{Im}}~t = -{\textrm{Im}}\int_0^r\frac{dr}{\sqrt{f(r)g(r)}}.
\label{1.27}
\end{eqnarray}
From the above one can easily show that ${\textrm{Im}}~t = -2\pi M$ for the Schwarzschild spacetime which is precisely the imaginary part of the transformation $t\rightarrow t-2i\pi M$ when one connects the two regions across the horizon as shown in \cite{Akhmedov}.  
Therefore the probability of the outgoing particle is
\begin{eqnarray}
P_{{\textrm{out}}}\sim{\textrm{exp}}\Big[-\frac{4}{\hbar}\omega\Big(1+\sum_i\beta_i\frac{\hbar^i}{M^{2i}}\Big){\textrm{Im}}\int_0^r\frac{dr}{\sqrt{f(r)g(r)}}\Big].
\label{1.28}
\end{eqnarray}
Now using the principle of ``detailed balance'' \cite{Paddy,Majhi2}
\begin{eqnarray}
P_{{\textrm{out}}}= {\textrm {exp}}\Big(-\frac{\omega}{T_h}\Big)P_{\textrm{in}}={\textrm{exp}} \Big(-\frac{\omega}{T_h}\Big)
\label{1.29}
\end{eqnarray}
we obtain the temperature of the black hole as
\begin{eqnarray}
T_h =T_H\Big(1+\sum_i\beta_i\frac{\hbar^i}{M^{2i}}\Big)^{-1}
\label{1.30}
\end{eqnarray}
where 
\begin{eqnarray}
T_H = \frac{\hbar}{4}\Big({\textrm{Im}}\int_0^r\frac{dr}{\sqrt{f(r)g(r)}}\Big)^{-1}
\label{1.31}
\end{eqnarray}
is the standard semiclassical Hawking temperature of the black hole and other terms are the corrections due to the quantum effect. Using this expression and knowing the metric coefficients $f(r)$ and $g(r)$ one can easily find out the temperature of the corresponding black hole. The same result was also obtained in \cite{Majhi3} for scalar particle tunneling.

      For the Schwarzschild black hole the metric coefficients are 
\begin{eqnarray}
f(r)=g(r)=(1-\frac{r_H}{r});\,\,\,r_H = 2M.
\label{1.32}
\end{eqnarray}
Therefore using (\ref{1.30}) and (\ref{1.31}) it is easy to write the corrected Hawking temperature:
\begin{eqnarray}
T_h=\frac{\hbar}{8\pi M}\Big(1+\sum_i\beta_i\frac{\hbar}{M^{2i}}\Big)^{-1}.
\label{1.33}
\end{eqnarray}
Now use of the Gibbs form of first law of thermodynamics gives the corrected form of the Bekenstein-Hawking entropy: 
\begin{eqnarray}
S_{\textrm{bh}}&=&\int\frac{dM}{T_h}= \frac{4\pi M^2}{\hbar}+8\pi \beta_1 \ln M - \frac{4\pi \hbar \beta_2}{M^2}+{\textrm{higher order terms in $\hbar$}}
\nonumber
\\
&=&\frac{\pi r_H^2}{\hbar}+8\pi\beta_1\ln r_H -\frac{16\pi\hbar\beta_2}{r_H^2}+{\textrm{higher order terms in $\hbar$}}
\label{1.34}
\end{eqnarray}
The area of the event horizon is
\begin{eqnarray}
A=4\pi r_H^2
\label{1.35}
\end{eqnarray}
so that, 
\begin{eqnarray}
S_{\textrm{bh}}&=&\frac{A}{4\hbar}+4\pi\beta_1\ln A-\frac{64\pi^2\hbar\beta_2}{A}+......................
\label{1.36}
\end{eqnarray}
It is noted that the first term is the usual semiclassical contribution to the area law $S_{\textrm{BH}}=\frac{A}{4\hbar}$ \cite{Bekenstein,Bardeen}. The other terms are the quantum corrections. Now it is possible to express the quantum corrections in terms of $S_{\textrm{BH}}$ by eliminating $A$:
\begin{eqnarray}
S_{\textrm{bh}}=S_{\textrm{BH}}+4\pi\beta_1\ln S_{\textrm{BH}}-\frac{16\pi^2\beta_2}{S_{\textrm{BH}}}+.........
\label{1.37}
\end{eqnarray}
Interestingly the leading order correction is logarithmic in $A$ or $S_{\textrm{BH}}$ which was found earlier in \cite{Fursaev,Mann} by field theory calculations and later in \cite{Majumdar,Mitra} by quantum geometry method. The higher order corrections involve inverse powers of $A$ or $S_{\textrm{BH}}$.

     To determine the value of the coefficients $\beta_1, \beta_2$ etc we will adopt the following steps. The point is that nonzero values for these coefficients are related to quantum corrections (loop effects). Such corrections, in a field theoretical approach, are manifested by the presence of anomalies. Now it is a well known fact that it is not possible to simultaneously preserve general coordinate (diffeomorphism) invariance and conformal invariance. Retaining general coordinate invariance, one finds the breakdown of conformal invariance leading to the presence of nonvanishing trace of the stress tensor.
We now show that the coefficients appearing in (\ref{1.37}) are related to this trace anomaly.

     We begin by studying the behaviour of the action (\ref{1.16}) upto order $\hbar^2$ under an infinitesimal constant scale transformation, parametrised by $k$, of the metric coefficients,
\begin{eqnarray}
\tilde{g}{_{\mu\nu}} = k g_{\mu\nu}\simeq(1+\delta k)g_{\mu\nu}.
\label{1.38}
\end{eqnarray}
Under this the metric coefficients of (\ref{1.01}) change as $\tilde f=kf,\tilde g=k^{-1}g$. Also, in order to preserve the scale invariance of the Dirac equation (\ref{1.02}), the field $\psi$ should transform as $\tilde\psi=k^{\frac{1}{2}}\psi$. On the other hand, $\psi$ has the dimension of ${(\textrm{mass})}^{\frac{3}{2}}$ and since in our case the only mass parameter is the black hole mass $M$, the infinitesimal change of it is given by,
\begin{eqnarray}
\tilde{M}=k^{\frac{1}{3}}M\simeq(1+\frac{1}{3}\delta k)M.
\label{1.39}
\end{eqnarray}

     Now from (\ref{1.28}) the imaginary part of the semiclassical contribution of the single particle action is
\begin{eqnarray}
\textrm{Im}I{_0}{_{(\textrm{out})}} = -2\omega{\textrm{Im}}~\int_0^r\frac{dr}{\sqrt{f(r)g(r)}}
\label{new1}
\end{eqnarray}
where $\omega$ gets identified with the energy (i.e. mass $M$) of a stable black hole \cite{Majhi1}. Therefore $\omega$  and $\hbar$ transforms like (\ref{1.39}) and $M^2$ respectively under (\ref{1.38}).

      Considering only the $\hbar$ and $\hbar^2$ order terms in (\ref{1.16}) and using (\ref{1.39}) we obtain, under the scale transformation,      
\begin{eqnarray}
{\tilde I}_{(1+2)}&\equiv&\hbar\textrm{Im}\tilde{I}{_1}{_{(\textrm{out})}}+\hbar^2\textrm{Im}\tilde{I}{_2}{_{(\textrm{out})}}
\nonumber
\\
&=&\Big(\frac{{\tilde\hbar}\beta_1}{\tilde{M}^2}+\frac{{\tilde{\hbar}}^2\beta_2}{\tilde{M}^4}\Big)\textrm{Im}\tilde I{_0}{_{(\textrm{out})}}
\nonumber
\\
&\simeq&\Big(\frac{\hbar\beta_1}{{M}^2}+\frac{{{\hbar}}^2\beta_2}{{M}^4}\Big)(1+\frac{1}{3}\delta k)\textrm{Im} I{_0}{_{(\textrm{out})}}
\nonumber
\\
&=&I_{(1+2)}+\Big(\frac{\hbar\beta_1}{{M}^2}+\frac{{{\hbar}}^2\beta_2}{{M}^4}\Big)\frac{1}{3}\delta k\textrm{Im} I{_0}{_{(\textrm{out})}}
\label{1.40}
\end{eqnarray} 
Therefore 
\begin{eqnarray}
\delta I_{(1+2)}&=&{\tilde{I}}_{(1+2)}-I_{(1+2)}
\nonumber
\\
&\simeq&\Big(\frac{\hbar\beta_1}{{M}^2}+\frac{{{\hbar}}^2\beta_2}{{M}^4}\Big)\frac{1}{3}\delta k\textrm{Im} I{_0}{_{(\textrm{out})}}
\label{ref}
\end{eqnarray}
leading to,
\begin{eqnarray}
\frac{\delta I_{(1+2)}}{\delta k}=\Big(\frac{\hbar\beta_1}{{M}^2}+\frac{{{\hbar}}^2\beta_2}{{M}^4}\Big)\frac{1}{3}\textrm{Im} I{_0}{_{(\textrm{out})}}
\label{1.41}
\end{eqnarray}
Now use of the definition of the energy-momentum tensor and (\ref{1.41}) yields,
\begin{eqnarray}
\textrm{Im}\int d^4x\sqrt{-g} T_\mu^\mu = \frac{2\delta I_{(1+2)}}{\delta k}
=\Big(\frac{\hbar\beta_1}{{M}^2}+\frac{{{\hbar}}^2\beta_2}{{M}^4}\Big)\frac{2}{3}\textrm{Im} I{_0}{_{(\textrm{out})}}
\label{1.42}
\end{eqnarray}
Thus, in the presence of a trace anomaly, the action is not invariant under the scale transformation. 
Since for the Schwarzschild black hole $f(r)$ and $g(r)$ are given by (\ref{1.32}), from (\ref{new1}) we obtain $\textrm{Im}S_0^{(\textrm{out})}=-4\pi\omega M$. Substituting this in (\ref{1.42}) we obtain for $\omega=M$ as,
\begin{eqnarray}
\hbar\beta_1+\frac{\hbar^2\beta_2}{M^2}= -\frac{3}{8\pi}\textrm{Im}\int d^4x\sqrt{-g}T_\mu^\mu
\label{1.43}
\end{eqnarray}
where $T_\mu^\mu$ is calculated upto two loops. Starting from the action (\ref{1.16}) and following the identical steps as above, a similer relation among all $\beta$'s with the right hand side of (\ref{1.43}) can be established. In this case $T_\mu^\mu$ is due to all loop expansions.

    Since the higher loop calculations to get $T_{\mu\nu}$ (from which $T_\mu^\mu$ is obtained) is very much complicated, usually in literature \cite{Witt} only one loop calculation for $T_{\mu\nu}$ is discussed. Thus, comparing only the $\hbar^1$ order on both sides of (\ref{1.43}), we obtain,
\begin{eqnarray}
\beta_1=-\frac{3}{8\pi}{\textrm{Im}}\int d^4x\sqrt{-g}{T^\mu_\mu}^{(1)}
\label{ref9}
\end{eqnarray}
This relation clearly shows that $\beta_1$ is connected to the trace anomaly.
A similar relation is given in \cite{Majhi4} where it has been shown that the coefficient $\beta_1$ is related to trace anomaly for the scalar particle tunneling. The only difference is the factor before the integration. This agrees well with the earlier conclusion \cite{Duff,Fursaev} where using conformal field theory technique it was shown that this $\beta_1$ is related to trace anomaly and is given by,
\begin{eqnarray}
\beta_1=-\frac{1}{360\pi}\Big(-N_0-\frac{7}{4}N_{\frac{1}{2}}+13N_1+\frac{233}{4}N_{\frac{3}{2}}-212N_2\Big)
\label{1.44}
\end{eqnarray}
`$N_s$' denotes the number of fields with spin `$s$'. In our case $N_{\frac{1}{2}}=1$ and $N_0=N_1=N_{\frac{3}{2}}=N_2=0$.\\\\

      To conclude, we have successfully extended our approach \cite{Majhi3} of scalar particle tunneling beyond semiclassical approximation to the model of fermion tunneling. We have considered all orders in the single particle action for fermion tunneling through the event horizon of the black hole. We showed that higher order correction terms of the action are proportional to the semiclassical contribution. A similar result was shown earlier in \cite{Majhi3} for the scalar particle tunneling. By dimensional argument and principle of ``detailed balance'' the same form of the modified Hawking temperature, as in the scalar case, was recovered. The logarithmic and inverse powers of area corrections to the Bekenstein-Hawking area law were also reproduced. Finally, we showed that the coefficient of the logarithmic term of entropy is related to trace anomaly. However, the prefactor appearing in this term is different from the scalar particle example, a result that is supported by earlier works \cite{Duff,Fursaev}.

     Here we have only told about $\beta_1$. Discussions on other coefficients can also be given from ({\ref{1.43}}), but since no information about $T_{\mu\nu}$ due to multi-loops is available in the literature, we cannot say anything about them at this moment. 
\\\\\\
{\it{Acknowledgment}}:\\
I wish to thank Prof. Rabin Banerjee for suggesting this investigation and constant encouragement.

\end{document}